\documentclass[12pt]{article}

\usepackage{latexsym,amsmath,amssymb,theorem,epsfig}

\DeclareFontFamily{OT1}{rsfs10}{}
\DeclareFontShape{OT1}{rsfs10}{m}{n}{ <-> rsfs10 }{}
\DeclareMathAlphabet{\mathscript}{OT1}{rsfs10}{m}{n}

\numberwithin{equation}{section}

\newcommand{\ns}{\normalsize}

\theoremstyle{plain}

{\theorembodyfont{\rmfamily} }

\begin{document}


\begin{titlepage}

\vspace{-5cm}

\title{
   \hfill{\ns UPR993-T} \\[1em]
   {\LARGE Superpotentials for Vector Bundle Moduli}
\\
[1em] }
\author{
   Evgeny I. Buchbinder$^1$, Ron Donagi$^2$ and
Burt A.~Ovrut$^1$ \\[0.5em]
   {\ns $^1$Department of Physics, University of Pennsylvania} \\[-0.4em]
   {\ns Philadelphia, PA 19104--6396}\\
   {\ns $^2$Department of Mathematics, University of Pennsylvania}
   \\[-0.4em]
   {\ns Philadelphia, PA 19104--6395, USA}\\}

\date{}

\maketitle

\begin{abstract}

We present a method for explicitly computing the non-perturbative
superpotentials associated with the vector bundle moduli in heterotic
superstrings
and $M$-theory. This method is applicable to any stable,
holomorphic vector bundle over an elliptically fibered Calabi-Yau
threefold.
For specificity, the vector bundle moduli superpotential, for a vector
bundle with
structure group $G=SU(3)$, generated by a heterotic
superstring wrapped once over an isolated curve in a Calabi-Yau threefold
with
base $B={\mathbb F}_{1}$, is explicitly calculated. Its locus of critical
points is discussed. Superpotentials of vector bundle moduli potentially
have
important implications for small instanton phase transitions and the
vacuum
stability and cosmology of superstrings and $M$-theory.

\end{abstract}

\thispagestyle{empty}

\end{titlepage}


\section{Introduction:}

The calculation of non-perturbative superpotentials for the moduli of
superstrings and $M$-theory has a considerable literature. The first
computations
were
carried out from the point of view of string worldsheet conformal field
theory \cite{DSWW1,DSWW2}.
Subsequently, a second approach appeared, pioneered in \cite{Witten1,
BBS},
in which
the associated worldsheet instantons are viewed as genus-zero holomorphic
curves $C$ in the compactification space, and one integrates over their
physical oscillations. This latter technique has been used to compute
non-perturbative superpotentials in $F$-theory \cite{DGW},
weakly coupled heterotic string theory on
Calabi-Yau manifolds \cite{Witten2}, $M$-theory compactified on
seven-manifolds of $G_{2}$
holonomy \cite{HM},
and heterotic $M$-theory on
Calabi-Yau threefolds
\cite{Lima1,Lima2}.
The results for
both the weakly and strongly coupled heterotic string theories are
proportional to a factor involving the Wess-Zumino term, which couples the
superstring to the background $SO(32)$ or $E_{8} \times E_{8}$
gauge bundle $V$
\cite{Witten2,Lima1,Lima2}.
This term can be expressed as the Pfaffian of a Dirac operator twisted by
the gauge bundle restricted to the associated holomorphic curve $C$.
It was pointed out in \cite{Witten2} that this Pfaffian, and, hence,
the superpotential,
will vanish if and only if the restriction of the gauge bundle,
$V|_{C}$, is non-trivial. Furthermore, it is clear that the Pfaffian
must be a holomorphic function of the gauge bundle moduli associated with
$V|_{C}$. Although related work has appeared in other contexts \cite{DGW},
neither the vanishing structure of the Pfaffian in heterotic string
theories,
nor its functional dependence on the vector bundle moduli, has yet
appeared in
the literature. It is the purpose of this paper to provide explicit
solutions
to these two problems, within the framework of both weakly and strongly
coupled
heterotic $E_{8} \times E_{8}$ superstring theories compactified on
elliptically fibered Calabi-Yau threefolds.

Our approach to determining the zeros of the Pfaffian is the following.
First, we note that the Pfaffian will vanish if and only if the chiral
Dirac
operator on the holomorphic curve $C$, in the background of the
restricted gauge bundle $V|_C$, has at least one zero mode. Thus,
the problem becomes one of determining whether or not the dimension of the
kernel of the Dirac operator is non-vanishing. We then show that this
kernel
naturally lies in a specific exact sequence of cohomologies and will be
non-vanishing if and only if the determinant of one of the maps in this
sequence vanishes. For a wide range of holomorphic vector bundles on
elliptically fibered Calabi-Yau threefolds, we can explicitly compute this
determinant as a holomorphic, homogeneous polynomial of the vector bundle
moduli associated with $V|_{C}$.
These moduli
parameterize a quotient manifold which is the projective
space of ``transition'' moduli introduced and described in~\cite{BDOold}.
It is then straightforward to determine its zeros and,
hence, the zeros of the Pfaffian.

It follows from this that the vanishing structure of the Pfaffian is
determined by a holomorphic polynomial function on the space of vector
bundle
moduli.
Note, however, that the Pfaffian must itself be a holomorphic function of
the
same moduli, and that this function must vanish at exactly the same locus
as does
the polynomial. Since the moduli space  is compact, one can conclude that
the
Pfaffian is given precisely by the holomorphic polynomial function,
perhaps to some
positive power, multiplied by an over-all constant. Using the results of
\cite{Bismut1,Bismut2,Bismut3}, one can show that this power must be unity.
Therefore, solving the
first problem, that is, the zeros of the Pfaffian, automatically solves
the second
problem, namely, explicitly determining the Pfaffian, and, hence, the
superpotential, as a function of the vector bundle moduli.

Specifically, in this paper we do the following. In Section 2 we discuss
the
definition of the Pfaffian, Pfaff$({\cal{D}}_{-})$, and its relation to
the
non-perturbative superpotential $W$.  We also show that $W$ is both gauge
and
local
Lorentz anomaly free. We begin our discussion of the structure of the
zeros
of Pfaff$({\cal{D}}_{-})$ in Section 3. This is accomplished in several
steps.
First, we show that the Pfaffian will vanish if and only if the dimension
of
the space of sections of a specific bundle on $C$, $h^{0}(C,
V|_{C} \otimes {\cal{O}}_{C}(-1))$, is non-vanishing. This condition is
shown to be identical to the non-triviality of $V|_{C}$ discussed in
\cite{Witten2}. As a second step,
we briefly review salient properties of elliptically
fibered Calabi-Yau threefolds and stable, holomorphic vector bundles over
such
spaces. In this paper, for specificity, we will restrict our discussion to
a
single, non-trivial example, with the base of the Calabi-Yau threefold
chosen
to be a Hirzebruch surface ${\mathbb F}_{1}$ and the vector bundle $V$
taken
to arise from an irreducible, positive spectral cover with structure group
$G=SU(3)$. Thirdly, we choose an isolated curve of genus zero,
${\cal{S}}={\mathbb P}^{1}$, in the Calabi-Yau threefold and wrap an
$E_{8} \times E_{8}$ heterotic superstring once around it. We explicitly
categorize $V|_{{\cal{S}}}$ and its associated moduli in terms of the
images on
${\cal{S}}$ of the spectral data. It is shown, using results
from \cite{BDOold},
that in our example these moduli parameterize a projective
space of dimension twelve,
${\mathbb P}^{12}$, which is the projectivization of the
thirteen-dimensional
linear
space of transition moduli. As the fourth and most important step, we
derive a
simple exact cohomology sequence in which the linear space of zero modes
of
the Dirac operator on ${\cal{S}}$, $H^{0}({\cal{S}}, V|_{{\cal{S}}}
\otimes
{\cal{O}}_{{\cal{S}}}(-1))$, appears. From this, we establish that its
dimension,
$h^{0}({\cal{S}}, V|_{{\cal{S}}} \otimes {\cal{O}}_{{\cal{S}}}(-1))$, will
be
non-vanishing if and only if the determinant of a certain linear map,
$f_{D}$,
in the exact sequence vanishes. As a last step, we explicitly compute
$detf_{D}$, finding it to be a polynomial of
degree twenty which exactly factors into the fourth power of a
holomorphic,
homogeneous polynomial of degree five in the thirteen homogeneous,
projective
moduli of $V|_{{\cal{S}}}$.
This solves the problem
of the vanishing structure of Pfaff$({\cal{D}}_{-})$.
In the final section, Section 4, we then argue that the
superpotential $W$ must be proportional to this degree twenty polynomial, 
thereby uniquely determining the
superpotential for the vector bundle moduli associated with
$V|_{{\cal{S}}}$.

As stated above, for simplicity, we have presented our results in terms of
a
single, non-trivial example. We have also suppressed much of the relevant
mathematics, emphasizing motivation and method over mathematical detail.
Our method, however, is, in principle,
applicable to any stable, holomorphic vector bundle
over
any elliptically fibered Calabi-Yau threefold.
In \cite{BDOnew}, we will present a wider
range of examples, computing the superpotentials for several different
vector
bundles and analyzing the structure of their critical points. In addition,
we
will
give a more complete discussion of the mathematical structure underlying
our
computations. Among other things, an analytical calculation
and first principles explanation of the homogeneous polynomials that occur
in vector bundle moduli superpotentials of the
type under discussion will be presented.

Although at first sight rather complicated to derive,
the superpotential for vector
bundle
moduli potentially has a number of important physical applications. To
begin
with, it is essential to the study of the stability of the vacuum
structure~\cite{CK,MPS} of
both weakly coupled heterotic string theory and
heterotic $M$-theory~\cite{hmt1,hmt2,hmt3}.
Furthermore, in both theories it allows, for the first time, a discussion
of
the dynamics of the gauge bundles. For example, in heterotic $M$-theory
one
can determine if a bundle is stable or whether it decays, via a small
instanton transition \cite{Pantev}, into five-branes.
In recent years, there has been
considerable research into the cosmology of superstrings and heterotic
$M$-theory \cite{low4, hos, blo} .
In particular, a completely new approach to early universe
cosmology, Ekpyrotic theory \cite{ekp1,ekp2,ekp3,ekp4,ekp5},
has been introduced within the context of
brane universe theories. The vector bundle superpotentials discussed in
this
paper and \cite{BDOnew} allow one to study the dynamics of the small
instanton phase
transitions that occur when a five-brane \cite{ekp1,ekp2} or
an "end-of-the-world" orbifold
plane \cite{ekp3,ekp4,ekp5} collides with our observable brane,
thus producing the Big Bang. These physical applications will be discussed
elsewhere.



\section{Pfaff$({\cal{D}}_{-})$ and Superpotential $W$:}


We want to consider $E_8 \times E_8$ heterotic superstring theory on the
space
\begin{equation}
M={\mathbb R}^4 \times X,
\label{1}
\end{equation}
where $X$ is a Calabi-Yau threefold. In general, this vacuum will admit a
stable, holomorhic vector bundle $V$ on $X$ with structure group
\begin{equation}
G \subseteq E_8 \times E_8
\label{2}
\end{equation}
and a specific connection one-form ${\cal{A}}$. It was shown in
\cite{CHSW} (see also \cite{GSW}) that for any open neighborhood of $X$,
the
local representative, $A$, of this connection satisfies the hermitian
Yang-Mills equations
\begin{equation}
F_{mn} = F_{\bar m \bar n} =0
\label{check}
\end{equation}
and
\begin{equation}
g^{m \bar n}F_{m \bar n} =0,
\label{3}
\end{equation}
where $F$ is the field strength of $A$.

As discussed in \cite{BBS}, a non-perturbative contribution to the
superpotential corresponds to the partition function of a superstring
wrapped
on a holomorphic curve $C \subset X$. Furthermore, one can show
\cite{Witten1} that only a curve of genus zero will contribute.
Hence, we will take
\begin{equation}
C={\mathbb P}^1.
\label{4.5}
\end{equation}
To further simplify the calculations, we will also assume that $C$
is an isolated curve in $X$ and that the superstring is wrapped only once
on $C$. The spin bundle over $C$ will be denoted by
\begin{equation}
S = S_{+} \oplus S_{-}
\label{5}
\end{equation}
and the restriction of the vector bundle $V$ to $C$ by $V|_{C}$.
Finally, we will assume that the structure group of the
holomorphic vector bundle is contained in the
subgroup $SO(16) \times SO(16)$
of $E_8 \times E_8$. That is,
\begin{equation}
G \subseteq SO(16) \times SO(16) \subset E_8 \times E_8.
\label{4}
\end{equation}
This condition will be satisfied by any quasi-realistic heterotic
superstring vacuum. Briefly, the reason for this restriction is the
following. As discussed, for example, in~\cite{Witten1,Lima1} and
references
therein, when~\eqref{4} is
satisfied, the Wess-Zumino-Witten ($WZW$) term coupling the superstring to
the
background vector bundle can be written as a theory of thirty-two
worldsheet
fermions interacting only with the vector bundle through the
covariant derivative. In this case, the associated partition function and,
hence, the contribution of the $WZW$ term to the superpotential is easily
evaluated. When condition~\eqref{4} is not satisfied, this procedure
breaks
down and the contribution of the $WZW$ term to the superpotential is
unknown.

Under these conditions, it can be shown \cite{Witten2} that the
non-perturbative
superpotential $W$ has the following structure
\begin{equation}
W  \propto {\rm Pfaff}({\cal D}_{-}) {\rm exp}({{\rm i} \int_{C} B}),
\label{star}
\end{equation}
where $B$ is the Neveu-Schwarz two-form field. The Pfaffian of
${\cal D}_{-}$ is defined as
\begin{equation}
{\rm Pfaff}({\cal D}_{-}) = \sqrt{det{\cal D}_{-}}
\label{6}
\end{equation}
where, for the appropriate choice of basis of the Clifford algebra,
\begin{equation}
{\cal D}_{-} = \bordermatrix{     & {\ } & {\ }  \cr
                             {\ } & 0 & D_{-} \cr
                             {\ } & i\partial_{+} & 0 \cr}.
\label{7}
\end{equation}
Here, the operator $D_{-}$ represents the covariant chiral Dirac operator
\begin{equation}
D_{-} : \Gamma (C, V|_{C} \otimes S_{-})
\rightarrow  \Gamma (C, V|_{C} \otimes S_{+}),
\label{cross}
\end{equation}
whereas $\partial_{+}: \Gamma (C, V|_{C} \otimes S_{+}) \rightarrow
\Gamma (C, V|_{C} \otimes S_{-})$ is independent of the connection
${\cal{A}}$. Pfaff$({\cal D}_{-})$ arises as the partition function of the
$WZW$ term, as discussed above.
Note that we have displayed in~\eqref{star} only those factors in the
superpotential relevant to vector bundle moduli. The factors omitted, such
as ${\rm exp} (\frac{- {\cal A} (C)}{2\pi \alpha^{\prime}})$
where ${\cal A}(C)$ is the area of the surface $C$ using the heterotic
string
Kahler metric on $X$ and $\alpha^{\prime}$ is the heterotic string
parameter,
are positive terms dependent on geometric moduli only.
Now
\begin{equation}
|det {\cal D}_{-}|^2 = det({\cal D}_{-} {\cal D}_{-}^{\dagger})
\propto det {\cal D},
\label{8}
\end{equation}
where the proportionality is a positive constant independent of the
connection,
\begin{equation}
{\cal D} = \bordermatrix{     & {\ } & {\ }  \cr
                             {\ } & 0 & D_{-} \cr
                             {\ } & D_{+} & 0 \cr}
\label{9}
\end{equation}
and
\begin{equation}
D_+ = D_{-}^{\dagger}.
\label{10}
\end{equation}
Note that we have absorbed a factor of $i$ into our definition of the
Dirac
operators $D_{-}$ and $D_{+}$.
It follows that
\begin{equation}
det {\cal D}_{-} \propto \sqrt{ |det {\cal D}|} {\rm e}^{{\rm i} \phi},
\label{11}
\end{equation}
where
\begin{equation}
|det{\cal D}| = detD_{-}D_{+}
\label{twostars}
\end{equation}
is a non-negative real number and $\phi$ is a phase. It is well known that
$det{\cal D}$ is gauge invariant. However, under both gauge and local
Lorentz transformations with infinitesimal parameters $\epsilon$ and
$\theta$ respectively, the phase can be shown to transform as
\begin{equation}
\delta \phi = 2 \int_C (-{\rm tr} (\epsilon d{\cal{A}}) + {\rm tr}
(\theta d\omega )),
\label{12}
\end{equation}
where ${\cal{A}}$ and $\omega$ are the gauge and spin connections
respectively.
This corresponds to the worldsheet sigma model anomaly. Fortunately, this
anomaly
is exactly cancelled by the variation
\begin{equation}
\delta B = \int_C ({\rm tr} (\epsilon d{\cal{A}}) -{\rm tr}
(\theta d\omega))
\label{13}
\end{equation}
of the $B$-field~\cite{Witten2}. It then follows from~\eqref{star}
that the
superpotential $W$ is both gauge and locally Lorentz invariant.

We displayed the factor ${\rm exp}({\rm i} \int_{C} B)$ in the
superpotential expression~\eqref{star} since it was relevant to the
discussion
of a gauge invariance. However, as was the case with
${\rm exp} (\frac{- {\cal A}(C)}{2\pi \alpha^{\prime}})$,
it also does not depend on the vector bundle moduli and, henceforth, we
will
ignore it. Therefore, to compute the vector bundle moduli contribution to
the superpotential one need only consider
\begin{equation}
W \propto {\rm Pfaff}({\cal D}_{-}).
\label{14}
\end{equation}
We now turn to the explicit calculation of ${\rm Pfaff}({\cal D}_{-})$. To
accomplish this, it is necessary first to discuss the conditions under
which it vanishes.


\section{The Zeros Of Pfaff$({\cal D}_{-})$:}


Clearly, Pfaff$({\cal D}_{-})$ vanishes if and only if $det{\cal D}$
does. In turn, it follows from~\eqref{twostars} that this will be the case
if and only if one or both of $D_{-}$ and $D_{+}$ have a non-trivial zero
mode.
In general, $dim{\ }kerD_{-}$ and  $dim{\ }kerD_{+}$ may not be equal to
each
other and must be considered separately. However, in this calculation that
is
not the case, as we now show. Recall that
\begin{equation}
index D_{+} = dim {\ } ker D_{+} - dim {\ } ker D_{-}.
\label{15}
\end{equation}
Since $dim_{{\mathbb R}}C=2$, it follows from the Atiyah-Singer index
theorem that
\begin{equation}
index D_{+} = \frac{i}{2\pi} \int_{ C}{{\rm tr} {\cal{F}}},
\label{16}
\end{equation}
where ${\cal{F}}$ is the curvature two-form associated with connection
${\cal{A}}$
restricted to curve $C$. Since, in this paper, the structure group of $V$
is contained in the semi-simple group $E_{8} \times E_{8}$,
we see that ${\rm tr}{\cal{F}}$ vanishes.
Therefore,
\begin{equation}
index D_{+} =0
\label{17}
\end{equation}
and, hence
\begin{equation}
dim{\ } ker D_{+} = dim{\ } ker D_{-}.
\label{18}
\end{equation}
It follows that Pfaff$({\cal D}_{-})$ will vanish if and only if
\begin{equation}
dim{\ } ker D_{-} >0 .
\label{threelines}
\end{equation}
To proceed, therefore, we must compute the zero structure of $dim{\ } ker
D_{-}$.
This calculation is facilitated using the fact that a holomorphic vector
bundle with a hermitian structure admits a unique connection compatible
with
both the metric and the complex
structure (see for example~\cite{GH}). That is,
for a special choice
of gauge, one can always set
\begin{equation}
D_{-} = i\bar \partial
\label{doublecheck}
\end{equation}
where, for any open neighborhood ${\cal{U}} \subset C$ with coordinates
$z, \bar z$,
\begin{equation}
\bar \partial = \partial_{\bar z}.
\label{19}
\end{equation}
To prove this, note that, for an appropriate choice of complex structure
and
Dirac $\gamma$-matrices one can always set
\begin{equation}
D_{-} = i(\partial_{\bar z} + A_{\bar z})
\label{bar}
\end{equation}
in any ${\cal{U}} \subset  C$. Now it follows from~\eqref{check} that
the pullback of the gauge connection on $X$ to any open set
${\cal{U}} \subset C$ is of the form
\begin{equation}
A_{z} = \partial_{z} g \cdot g^{-1},
\qquad A_{\bar z} = \partial_{\bar z} g^{\dagger -1}\cdot
g^{\dagger}
\label{21}
\end{equation}
where $g$ is a map from ${\cal{U}}$ to the complexification of structure
group $G$.
Now choose
a specific open neighborhood ${\cal{U}} \subset C$ with local coordinates
$z, \bar z$. Clearly, using a gauge transformation with parameter
$g^{\dagger}$ we can set
\begin{equation}
A_{\bar z} =0
\label{triplecheck}
\end{equation}
in this ${\cal{U}}$. Now let ${\cal{V}}$ be any other open subset of $C$
with local
coordinates $z^{\prime}, \bar z^{\prime}$ , such that it has a non-empty
intersection with ${\cal{U}}$. Denote the connection one-form on
${\cal{V}}$ by
$A^{\prime}$.
Then, the compatibility condition says that
\begin{equation}
A^{\prime} = f_{{\cal{U}}{\cal{V}}}Af^{-1}_{{\cal{U}}{\cal{V}}} +
df_{{\cal{U}}{\cal{V}}} \cdot f^{-1}_{{\cal{U}}{\cal{V}}},
\label{22}
\end{equation}
where $f_{{\cal{U}}{\cal{V}}}$ is the holomorphic transition fuction on
${\cal{U}} \cap {\cal{V}}$.
It follows from equation~\eqref{triplecheck} and from the fact that
$\partial_{\bar z}f_{{\cal{U}}{\cal{V}}}=0$ that
\begin{equation}
A^{\prime}_{\bar z^{\prime}}=0.
\label{23}
\end{equation}
Continuing this way we see that one can set $A_{\bar z}$ to zero globally
on $C$. It then follows from relation~\eqref{bar} that
$D_{-} = \bar \partial$ as claimed in~\eqref{doublecheck}. Note, in
passing,
that with respect to the same complex structure and in the same gauge,
the operator $D_{+}$ is of the form
\begin{equation}
D_{+} = i(\partial_{z} + A_{z})
\label{24}
\end{equation}
where, in general, $A_{z} \neq 0$ on every open subset of $C$.
Using~\eqref{threelines} and~\eqref{doublecheck}, we conclude that
Pfaff$({\cal D}_{-})$ will vanish if and only if
\begin{equation}
dim{\ } ker \bar \partial >0.
\label{25}
\end{equation}
However, it follows from equations~\eqref{cross} and~\eqref{doublecheck}
that
the zero modes of $\bar \partial$ are precisely the holomorphic sections
of the vector bundle $V|_{C} \otimes S_{-}$. Using the fact that
on $C={\mathbb P}^1$
\begin{equation}
S_{-} ={\cal O}_{C}(-1),
\label{26}
\end{equation}
and defining
\begin{equation}
V|_C (-1) =  V|_{C} \otimes {\cal O}_{C}(-1),
\label{27}
\end{equation}
we conclude that
\begin{equation}
dim{\ } ker \bar \partial = h^{0} (C, V|_{C} (-1)).
\label{28}
\end{equation}
Hence, Pfaff$({\cal D}_{-})$ will vanish if and only if
\begin{equation}
h^{0} (C, V|_{C} (-1))>0.
\label{29}
\end{equation}
Therefore, the problem of determining the zeros of the Pfaffian
of ${\cal D}_{-}$ is reduced to deciding whether or not there are
any
non-trivial global holomorphic sections of the bundle $V|_{C}(-1)$ over
the
curve $C$. An equivalent way of stating the same result is to realize that
the condition for the vanishing of Pfaff$({\cal D}_{-})$ is directly
related
to the non-triviality or triviality of the bundle $V|_{C}$. To see this,
note
that any holomorphic $SO(16) \times SO(16)$ bundle $V|_{C}$ over a genus
zero
curve $C ={\mathbb P}^{1}$ is of the form
\begin{equation}
V|_{C} = \bigoplus_{i=1}^{16} {\cal O}_{{\mathbb P}^{1}}(m_i) \oplus
{\cal O}_{{\mathbb P}^{1}}(-m_i)
\label{30}
\end{equation}
with non-negative integers $m_i$. Therefore
\begin{equation}
V|_{C}(-1) = \bigoplus_{i=1}^{16} {\cal O}_{{\mathbb P}^{1}}(m_i-1)
\oplus {\cal O}_{{\mathbb P}^{1}}(-m_i-1).
\label{31}
\end{equation}
Using the fact that
\begin{equation}
h^{0}({\mathbb P}^1, {\cal O}_{{\mathbb P}^1} (m)) = m+1
\label{alpha}
\end{equation}
for $m \geq 0$ and
\begin{equation}
h^{0}({\mathbb P}^1, {\cal O}_{{\mathbb P}^1} (m)) = 0
\label{beta}
\end{equation}
for $m < 0$, it follows that
\begin{equation}
h^{0} (C, V|_{C} (-1))= \sum_{i=1}^{16} m_i.
\label{32}
\end{equation}
Therefore  $h^{0} (C, V|_{C} (-1))>0$ if and only if at least one $m_i$
is greater than zero. That is, as first pointed out in
\cite{Witten2},
$h^{0} (C, V|_{C} (-1))>0$, and hence Pfaff$({\cal D}_{-})$ will vanish,
if and only if $V|_{C}$ is non-trivial.
We now turn to the question of how to determine whether or not there are
non-trivial sections of $V|_{C}(-1)$ over $C$.

The problem of whether or not $h^{0} (C, V|_{C} (-1))$ is non-zero
can be solved within the context of stable, holomorphic vector bundles
over
elliptically fibered Calabi-Yau threefolds. In this paper, we will present
a single explicit example, prefering to be concrete and to emphasize the
method rather than the underlying mathematics. A more detailed discussion,
with all the relevant mathematics, will be presented elsewhere
\cite{BDOnew}.
We consider a Calabi-Yau threefold $X$ elliptically fibered over a base
\begin{equation}
B={\mathbb F}_1,
\label{33}
\end{equation}
where ${\mathbb F}_1$ is a Hirzebruch surface. That is,
$\pi:X \to {\mathbb F}_1$. Since $X$ is elliptically fibered, there exists
a zero section $\sigma : {\mathbb F}_1 \to X$. We will denote
$\sigma({\mathbb F}_1) \subset X$ simply as $\sigma$.
The second homology group $H^{2}({\mathbb F}_1, {\mathbb R})$ is spanned
by two effective classes of curves, denoted by ${\cal S}$ and ${\cal E}$,
with
intersection numbers
\begin{equation}
{\cal S}^2 = -1, \qquad {\cal S} \cdot {\cal E} = 1, \qquad
{\cal E}^2 =0.
\label{aaa}
\end{equation}
The first Chern class of ${\mathbb F}_1$ is given by
\begin{equation}
c_1 ({\mathbb F}_1) = 2{\cal S} + 3{\cal E}.
\label{A}
\end{equation}
Over $X$ we construct a stable, holomorphic vector bundle $V$ with
structure
group
\begin{equation}
G=SU(3).
\label{34}
\end{equation}
This is accomplished \cite{FMW,DO} by specifying a spectral cover
\begin{equation}
{\cal C} = 3 \sigma + \pi^{*} \eta ,
\label{35}
\end{equation}
where
\begin{equation}
\eta = (a+1) {\cal S} + b{\cal E}
\label{B}
\end{equation}
and $a+1$ and $b$ are non-negative integers, as well as a holomorphic line
bundle
\begin{equation}
{\cal N} = {\cal O}_{X} ( 3(\lambda +\frac{1}{2}) \sigma  -
(\lambda -\frac{1}{2}) \pi^{*}\eta + (3 \lambda +\frac{1}{2})
\pi^{*}c_1(B)),
\label{C}
\end{equation}
where $\lambda \in {\mathbb Z}+\frac{1}{2}$.
Note that we use $a+1$, rather than $a$, as the coefficient of ${\cal{S}}$
in~\eqref{B} to conform with our conventions in~\cite{BDOold}.
The vector bundle $V$ is then determined via a Fourier-Mukai
transformation
\begin{equation}
({\cal C}, {\cal N}) \longleftrightarrow V.
\label{37}
\end{equation}
In this paper, we will consider the case
\begin{equation}
a > 5, \qquad b-a =6, \qquad \lambda =\frac{3}{2}.
\label{D}
\end{equation}
We refer the reader to \cite{BDOold} to show that for such parameters
spectral cover ${\cal C}$ is both irreducible and positive. In addition,
it follows from~\eqref{A},~\eqref{B},~\eqref{C} and~\eqref{D}
that
\begin{equation}
{\cal N} = {\cal O}_{X} (6 \sigma  +
(9-a)\pi^{*}({\cal S} + {\cal E})).
\label{38}
\end{equation}
Now consider the curve ${\cal S} \subset {\mathbb F}_1$. Since
${\cal S} \cdot {\cal S} =-1$, it is an isolated curve in ${\mathbb F}_1$.
Since ${\cal S}$ is an exceptional curve
\begin{equation}
{\cal S} ={\mathbb P}^{1}.
\label{39}
\end{equation}
The lift of ${\cal S}$ into $X$, $\pi^{*} {\cal S}$, was determined
in \cite{BDOold} to be the rational elliptic surface
\begin{equation}
\pi^{*}{\cal S} =dP_{9}.
\label{40}
\end{equation}
The curve ${\cal S}$ is represented in $X$ by
\begin{equation}
{\cal S}_{X} =\sigma \cdot \pi^{*}{\cal S}.
\label{41}
\end{equation}
By construction, ${\cal S}_{X}$ is isolated in $X$ and
${\cal S}_{X} ={\mathbb P}^{1}$. We will frequently not distiguish between
${\cal S}$ and ${\cal S}_{X}$, referring to both curves as ${\cal S}$.
Recall that we want to wrap the superstring once over a genus-zero Riemann
surface
${\mathbb P}^{1}$ which is isolated in $X$. In this example, we will take
${\cal S}$ to be this isolated curve.

To proceed, let us restrict the vector bundle data to
$\pi^{*}{\cal S}$. The restriction of the spectral cover is given by
\begin{equation}
{\cal C}|_{dP_9} = {\cal C}\cdot \pi^{*}{\cal S}
\label{42}
\end{equation}
which, using~\eqref{aaa} and~\eqref{D}, becomes
\begin{equation}
{\cal C}|_{dP_9} = 3 \sigma|_{dP_9} + 5 F,
\label{43}
\end{equation}
where $F$ is the class of the elliptic fiber. Note that
${\cal C}|_{dP_9}$ is a divisor in $dP_9$. Similarly
\begin{equation}
{\cal N}|_{dP_9} = {\cal O}_{dP_9} ((6\sigma +(9-a)
\pi^{*} ({\cal S}+{\cal E}))\cdot \pi^{*}{\cal S}).
\label{44}
\end{equation}
Using~\eqref{aaa}, this is given by
\begin{equation}
{\cal N}|_{dP_9} = {\cal O}_{dP_9} (6\sigma|_{dP_9}).
\label{bbb}
\end{equation}
It is useful, as will be clear shortly, to define
\begin{equation}
{\cal N}|_{dP_9}(-F) ={\cal N}|_{dP_9} \otimes {\cal O}_{dP_9} (-F).
\label{ddd}
\end{equation}
Then
\begin{equation}
{\cal N}|_{dP_9}(-F) ={\cal O}_{dP_9} (6\sigma|_{dP_9}-F).
\label{eee}
\end{equation}
Since $dP_9$ is elliptically fibered, the restriction of $V$ to
$dP_9$, denoted by $V|_{dP_9}$, can be obtained from the Fourier-Mukai
transformation
\begin{equation}
({\cal C}|_{dP_9}, {\cal N}|_{dP_9}) \longleftrightarrow V|_{dP_9}.
\label{45}
\end{equation}

In a previous paper
\cite{BDOold}, we showed that the direct image under $\pi$ of the line
bundle
on $dP_9$ associated with ${\cal C}|_{dP_9}$, that is
\begin{equation}
{\cal O}_{dP_9}(3 \sigma|_{dP_9} + 5 F)
\label{46}
\end{equation}
is a rank three vector bundle on ${\cal S}$. In this case, we find that
\begin{equation}
\pi_{*} {\cal O}_{dP_9} (3 \sigma|_{dP_9} + 5F) =
{\cal O}_{{\cal S}}(5) \oplus {\cal O}_{{\cal S}}(3) \oplus
{\cal O}_{{\cal S}}(2).
\label{mod1}
\end{equation}
In addition, we proved in \cite{BDOold} that the moduli associated with
a small instanton phase transition involving the curve ${\cal S}$,
the so called transition moduli, are in one-to-one
correspondence with the holomorphic sections of this bundle, that is,
with elements of
\begin{equation}
H^{0}({\cal S}, {\cal O}_{{\cal S}}(5) \oplus {\cal O}_{{\cal S}}(3)
\oplus
{\cal O}_{{\cal S}}(2)).
\label{mod2}
\end{equation}
It follows that the number of these transition moduli is given by
\begin{equation}
h^{0}({\cal S}, {\cal O}_{{\cal S}}(5) \oplus {\cal O}_{{\cal S}}(3)
\oplus {\cal O}_{{\cal S}}(2)) =13,
\label{47}
\end{equation}
where we have used expression~\eqref{alpha}.
In this paper, we are interested not in the
vector space ~\eqref{mod2} parametrized by the
full set of transition moduli but,
rather, in its projectivization
\begin{equation}
{\mathbb P}^{12} \simeq  {\mathbb P}H^{0}(dP_9, {\cal O}_{dP_9}(3
\sigma|_{dP_9} +5F)),
\label{rock4}
\end{equation}
parametrized by
the moduli of
the curve
${\cal{C}}|_{dP_{9}}$.
This space, although twelve dimensional, is most easily parameterized in
terms
of the thirteen homogeneous coordinates $a_{i}$.
Now note that ${\cal C}|_{dP_9}$ is a $3$-fold cover of ${\cal S}$ with
covering map
$\pi_{{\cal C}|_{dP_9}}:{\cal C}|_{dP_9} \to {\cal S}$.
The image of ${\cal N}|_{{\cal C}|_{dP_9}}$
under $\pi_{{{\cal C}}|_{dP_9}}$ is also
a rank three vector bundle over ${\cal S}$. In fact,
\begin{equation}
V|_{{\cal S}} =\pi_{{\cal C}|_{dP_9}*}{\cal N}|_{{\cal{C}}|_{dP_9}}.
\label{ccc}
\end{equation}

Within this context, we can now consider the question of determining the
zeros
of Pfaff$({\cal D}_{-})$ for the explicit case of a superstring wrapped on
${\cal S}$. As discussed in the previous section, we want to study
the properties of $h^{0}({\cal S}, V|_{{\cal S}}(-1))$. Now,
using~\eqref{ccc}
we have
\begin{equation}
h^{0}({\cal S}, V|_{{\cal S}}(-1))=
h^{0}({\cal S}, \pi_{{\cal C}|_{dP_9}*}{\cal N}|_{dP_9}
\otimes {\cal O}_{{\cal S}}(-1)).
\label{burt}
\end{equation}
Using a Leray spectral sequence and~\eqref{ddd}, one can show that
\begin{equation}
h^{0}({\cal S}, \pi_{{\cal C}|_{dP_9}*}{\cal N}|_{dP_9}
\otimes {\cal O}_{{\cal S}}(-1))=
h^{0} ({\cal C}|_{dP_9}, {\cal N}|_{dP_9}(-F)|_{{\cal C}|_{dP_9}})
\label{evgeny}
\end{equation}
where, for our specific example, ${\cal N}|_{dP_9}(-F)$ is given
by~\eqref{eee}. In other words, we will find under what circumstances
the vector bundle $V|_{{\cal S}}(-1)$ has sections by studying under what
circumstances the line bundle ${\cal N}|_{dP_9}(-F)$ restricted
to ${\cal C}|_{dP_9}$ does. To accomplish this, consider the short exact
sequence
\begin{equation}
0 \rightarrow E \otimes {\cal O}_{dP_9} (-D) \stackrel{f_{D}}{\rightarrow}
E
\stackrel{r}{\rightarrow} E|_{D} \rightarrow 0,
\label{xxx}
\end{equation}
where $E$ is any holomorphic vector bundle on $dP_9$ and $D$ is any
effective
divisor in $dP_9$. The map $f_{D}$ from $E \otimes {\cal O}_{dP_9} (-D)$
to $E$ is given by multiplication by the section of the line bundle
${\cal O}_{dP_9} (D)$ that vanishes precisely on $D$. The mapping $r$
from $E$ to $E|_{D}$ is just restriction. If we use the abbreviation
\begin{equation}
E(-D)=E \otimes {\cal O}_{dP_9} (-D)
\label{yyyy}
\end{equation}
then, in the usual way,~\eqref{xxx} implies the long exact sequence of
cohomology
groups given by
\begin{eqnarray}
&&0 \rightarrow H^{0}(dP_9, E(-D))
\rightarrow
H^{0}(dP_9, E) \rightarrow
H^{0}(D, E|_{D}) \nonumber \\
&&\rightarrow H^{1}(dP_9, E(-D))
\rightarrow
H^{1}(dP_9, E))
\rightarrow
H^{1}(D, E|_{D}) \rightarrow \dots
\quad .
\label{big}
\end{eqnarray}
In our specific example, we choose
\begin{equation}
E= {\cal N}|_{dP_9} (-F) = {\cal O}_{dP_9} (6 \sigma|_{dP_9} -F),
\label{48}
\end{equation}
where we have used~\eqref{eee}, and
\begin{equation}
D={\cal C}|_{dP_9} = 3 \sigma|_{dP_9} + 5F.
\label{divisor}
\end{equation}
It follows from~\eqref{yyyy} that
\begin{equation}
E(-D)={\cal O}_{dP_9}(3 \sigma|_{dP_9} -6F).
\label{zzzz}
\end{equation}
Let us first first consider the term $H^{0}(dP_{9},E)=H^{0}(dP_9, {\cal
O}_{dP_9}
(6 \sigma|_{dP_9} -F))$ in the long exact sequence.
We compute it in terms of its direct image on ${\cal{S}}$:
\begin{equation}
\pi_{*}{\cal O}_{dP_9} (6 \sigma|_{dP_9} -F) =
{\cal O}_{{\cal{S}}}(-1) \oplus \bigoplus_{i=3}^{7} {\cal
O}_{{\cal{S}}}(-i).
\label{49}
\end{equation}
It follows from~\eqref{beta} that this bundle has no global holomorphic
sections and, hence, that
\begin{equation}
H^{0}(dP_9, E)=0.
\label{50}
\end{equation}
Furthermore, note from~\eqref{eee},~\eqref{48} and~\eqref{divisor} that
\begin{equation}
H^{0}(D, E|_{D}) =
H^{0}({\cal C}|_{dP_9}, {\cal N}|_{dP_9}
(-F)|_{{\cal C}|_{dP_9}}).
\label{51}
\end{equation}
Therefore, part of the exact sequence~\eqref{big} is given by
\begin{eqnarray}
&&0 \rightarrow H^{0}({\cal C}|_{dP_9}, {\cal N}|_{dP_9}
(-F)|_{{\cal C}|_{dP_9}})
\rightarrow H^{1}(dP_9, {\cal O}_{dP_9}(3 \sigma|_{dP_9} -6F))
\nonumber \\
&&\rightarrow
H^{1}(dP_9, {\cal O}_{dP_9}(6 \sigma|_{dP_9} -F))
\rightarrow \cdots \quad .
\label{exact}
\end{eqnarray}
Both $H^{1}$ cohomology groups are linear spaces whose structure and
dimension we now determine. First consider
$H^{1}(dP_9, {\cal O}_{dP_9}(3 \sigma|_{dP_9} -6F))$. Using a Leray
spectral
sequence and the facts that
\begin{equation}
\pi_{*} {\cal O}_{dP_9}(3 \sigma|_{dP_9} -6F) =
{\cal O}_{{\cal S}}(-6) \oplus {\cal O}_{{\cal S}}(-8)
\oplus {\cal O}_{{\cal S}}(-9)
\label{52}
\end{equation}
and
\begin{equation}
R^{1}\pi_{*} {\cal O}_{dP_9}(3 \sigma|_{dP_9} -6F) = 0,
\label{52a}
\end{equation}
it follows that
\begin{equation}
H^{1}(dP_9, {\cal O}_{dP_9}(3 \sigma|_{dP_9} -6F)) =
H^{1} ({\cal S}, {\cal O}_{{\cal S}}(-6) \oplus {\cal O}_{{\cal S}}(-8)
\oplus {\cal O}_{{\cal S}}(-9)).
\label{first}
\end{equation}
Furthemore, using the Serre duality
\begin{equation}
H^{1}({\mathbb P}^{1}, {\cal O}_{{\mathbb P}^{1}}(p)) =
H^{0}({\mathbb P}^{1}, {\cal O}_{{\mathbb P}^{1}}(-2-p))^{*},
\label{serre}
\end{equation}
where $p$ is any integer and $*$ signifies the dual linear space, we see
that
\begin{equation}
H^{1} ({\cal S}, {\cal O}_{{\cal S}}(-6) \oplus {\cal O}_{{\cal S}}(-8)
\oplus {\cal O}_{{\cal S}}(-9))=
H^{0} ({\cal S}, {\cal O}_{{\cal S}}(4) \oplus {\cal O}_{{\cal S}}(6)
\oplus {\cal O}_{{\cal S}}(7))^{*}.
\label{second}
\end{equation}
If we denote
\begin{equation}
W_1 = H^1 (dP_9, {\cal O}_{dP_9} (3 \sigma|_{dP_9} -6F)),
\label{w1}
\end{equation}
then it follows from~\eqref{first},~\eqref{second} and~\eqref{alpha} that
\begin{equation}
dim W_1 = 20.
\label{53}
\end{equation}
Now consider $H^{1}(dP_9, {\cal O}_{dP_9} (6 \sigma|_{dP_9} -F))$.
Exactly as above, one can show using a Leray spectral sequence that
\begin{equation}
H^1 (dP_9, {\cal O}_{dP_9} (6 \sigma|_{dP_9} -F))=
H^{1} ({\cal S}, {\cal O}_{{\cal S}}(-1) \oplus
\bigoplus_{i=3}^{7} {\cal O}_{{\cal S}}(-i))
\label{third}
\end{equation}
and, by Serre duality, that
\begin{equation}
H^{1} ({\cal S}, {\cal O}_{{\cal S}}(-1) \oplus
\bigoplus_{i=3}^{7} {\cal O}_{{\cal S}}(-i))
=
H^{0} ({\cal S}, {\cal O}_{{\cal S}}(-1) \oplus
\bigoplus_{i=3}^{7} {\cal O}_{{\cal S}}(-2+i))^{*}.
\label{fourth}
\end{equation}
If we denote
\begin{equation}
W_2 = H^{1} (dP_9, {\cal O}_{dP_9} (6 \sigma|_{dP_9} -F)),
\label{w2}
\end{equation}
then it follows from~\eqref{third},~\eqref{fourth} and~\eqref{alpha} that
\begin{equation}
dim W_2 =20
\label{54}
\end{equation}
as well.

Let us now reconsider the structure of the exact sequence~\eqref{exact}.
This can now be written as
\begin{equation}
0 \rightarrow H^{0}({\cal S}, V|_{\cal S}(-1))
\rightarrow W_1 \stackrel{f_D}{\rightarrow} W_2 \rightarrow \cdots
\label{ww}
\end{equation}
where we have used~\eqref{burt},~\eqref{evgeny},\eqref{w1} and~\eqref{w2}.
We will now display the linear mapping $f_D$. This is induced
from the map $f_D$ in the short exact sequence
\begin{equation}
0 \rightarrow E \otimes {\cal O}_{dP_9} (-D)
\stackrel{f_D}{\rightarrow} E
\rightarrow E|_{D} \rightarrow 0.
\label{55}
\end{equation}
As discussed previously, the map $f_D$ in~\eqref{55} is just
multiplication
by the unique, up to scaling, element of $H^{0}(dP_9,
{\cal O}_{dP_9} (D))$ with the
property that it vanishes on $D$. In the present example,
$D={\cal{C}}|_{dP_{9}}$.

Exact sequence~\eqref{ww} is precisely what we need to solve the problem
of whether or not $h^{0}({\cal S}, V|_{{\cal S}}(-1))$ is zero.
Since $W_1$ and $W_2$ are just linear spaces of the same dimension, and
since
it follows from~\eqref{ww}
that the space we are interested in, $H^{0}({\cal S}, V|_{{\cal S}}(-1))$,
is the kernel of the map $f_{{\cal{C}}|_{dP_{9}}}$, we conclude that
$h^{0}({\cal S}, V|_{{\cal S}}(-1))>0$ if and only if
\begin{equation}
det{\ }f_{{\cal{C}}|_{dP_{9}}} =0.
\label{58}
\end{equation}
Therefore, the solution to this problem, and hence to
finding the zeros of Pfaff$({\cal D}_{-})$,
reduces
to computing $det{\ }f_{{\cal{C}}|_{dP_{9}}}$, to which we now turn.
An arbitrary element of $W_1$ can be charactarized as follows. Let
\begin{equation}
\tilde{w}_1 =  B_{-6} \oplus B_{-8} \oplus B_{-9}
\label{59}
\end{equation}
be an element of $H^{1} ({\cal S}, {\cal O}_{{\cal S}} (-6) \oplus
{\cal O}_{{\cal S}} (-8) \oplus {\cal O}_{{\cal S}} (-9))$ where
$B_{ -i}$, $i=6, 8, 9$ denotes an arbitrary section in
$H^{1}({\cal S}, {\cal O}_{\cal S}(-i))$. We see from the Serre duality
relation~\eqref{serre} and~\eqref{alpha} that
\begin{equation}
h^{1}({\cal S}, {\cal O}_{\cal S}(-i)) = i-1.
\label{dim}
\end{equation}
Now let us lift $\tilde{w}_1$ to $w_1$ in
$H^{1} (dP_9, {\cal O}_{dP_9} (3 \sigma|_{dP_9}-6F))$,
using~\eqref{first}.
We find that
\begin{equation}
w_{1} = b_{-6} z + b_{-8} x + b_{-9} y,
\label{lift}
\end{equation}
where, from the isomorphism
\begin{equation}
{\cal O}_{dP_9} (kF) = \pi^{*} {\cal O}_{{\cal S}} (k)
\label{60}
\end{equation}
for any integer $k$,
$b_{-i}=\pi^{*}B_{-i}$ are elements in
$H^{1}(dP_9, {\cal O}_{dP_{9}}(-iF))$ and we have used the fact
that $dP_9$ has a Weierstrass representation
\begin{equation}
y^2 z=4 x^3 - g_2 x z^2 - g_3 z^3,
\label{61}
\end{equation}
where~\cite{BDOold}
\begin{equation}
x \sim {\cal O}_{dP_9} (3 \sigma|_{dP_9} +2F), \qquad
y \sim {\cal O}_{dP_9} (3 \sigma|_{dP_9} +3F), \qquad
z \sim {\cal O}_{dP_9} (3 \sigma|_{dP_9})
\label{coord}
\end{equation}
and
\begin{equation}
g_2 \sim {\cal O}_{dP_9} (4F), \qquad g_3 \sim {\cal O}_{dP_9} (6F).
\label{coord2}
\end{equation}
In the above equations, symbol $\sim$ means ``section of''.

Expression~\eqref{lift} completely characterizes an element
$w_1 \in W_1$. In a similar way, any element
$w_2 \in W_2= H^{1}(dP_9, {\cal O}_{dP_9} (6 \sigma|_{dP_9} -F))$
can be written
as
\begin{equation}
w_{2} = c_{-3} zx + c_{-4} zy + c_{-5} x^2 + c_{-6}xy + c_{-7} y^2
\label{lift2}
\end{equation}
where for $j=3, \cdots 7$, $c_{-j}=\pi^{*}C_{-j}$ is an element of
$H^{1}(dP_9, {\cal O}_{dP_9} (-jF))$ and $C_{-j}$ is a section in the
$j-1$-dimensional space $H^{1}({\cal{S}}, {\cal{O}}_{{\cal{S}}}(-j))$.
Equation~\eqref{lift2} follows from expressions~\eqref{third},~\eqref{60}
and~\eqref{coord}. Finally, we note from~\eqref{mod1} and~\eqref{coord} that
any map
$f_{{\cal{C}}|_{dP_{9}}}$ can be expressed as
\begin{equation}
f_{{\cal{C}}|_{dP_{9}}} = m_5 z + m_3 x + m_2 y,
\label{lift3}
\end{equation}
where $m_k=\pi^{*}M_{k}$, $k=2, 3, 5$, is an element in
$H^{0}(dP_9, {\cal O}_{dP_9} (kF))$ and $M_{k}$ is a section in the
$k+1$-dimensional space $H^{0}({\cal{S}},{\cal{O}}_{{\cal{S}}}(k))$.
Although
there are thirteen parameters in $m_{k}$, $k=2,3,5$, it must be remembered
that
they are homogeneous coordinates for the twelve dimensional projective
space
${\mathbb P}H^{0}(dP_9, {\cal O}_{dP_9} (3 \sigma|_{dP_9} +5F))$.

Putting this all together, we can completely specify the linear mapping
$W_1 \stackrel{f_{{\cal{C}}|_{dP_{9}}}}{\rightarrow} W_2$.
First note that with respect to fixed basis vectors of $W_{1}$ and
$W_{2}$,
the linear map $f_{{\cal{C}}|_{dP_{9}}}$ is a $20 \times 20$ matrix. In
order
to find this matrix explicitly, we have to study its action on these
vectors. This action is generated through
multiplcation by a section $f_{{\cal{C}}|_{dP_{9}}}$
of the form~\eqref{lift3}. Suppressing, for
the time being, the vector coefficients $b_{-i}$ and $c_{-j}$,
we see from~\eqref{lift} that the linear space $W_1$ is spanned by the
basis vector blocks
\begin{equation}
z, \quad x, \quad y
\label{62}
\end{equation}
whereas it follows from~\eqref{lift2} that the linear space $W_2$ is
spanned by basis vector blocks
\begin{equation}
zx, \quad zy, \quad x^2, \quad xy, \quad y^2.
\label{63}
\end{equation}
The explicit matrix $M_{IJ}$ representing
$f_{{\cal{C}}|_{dP_{9}}}$ is determined by multiplying the
basis vectors~\eqref{62} of $W_{1}$ by $f_{{\cal{C}}|_{dP_{9}}}$
in~\eqref{lift3}. Expanding the
resulting vectors in $W_{2}$ in the
basis~\eqref{63} yields the matrix.  We find that $M_{IJ}$ is given by
\begin{equation}
\bordermatrix{    & z   & x   & y  \cr
              xz  & m_3 & m_5 & 0  \cr
              yz  & m_2 & 0   & m_5 \cr
              x^2 & 0   & m_3 & 0   \cr
              xy  & 0   & m_2 & m_3 \cr
              y^2 & 0   & 0   & m_2 \cr}.
\label{table}
\end{equation}
Of course, $M_{IJ}$ is a $20 \times 20$ matrix, so each of the elements
of~\eqref{table} represents
a $(j-1) \times (i-1)$ matrix for the corresponding $j=3, 4, 5, 6, 7$
and $i=6, 8, 9$.
For example, let us compute $M_{11}$. This
corresponds to the $xz - z$ component of~\eqref{table} where
\begin{equation}
H^{1} (dP_9, {\cal O}_{dP_9} (3 \sigma|_{dP_9} -6F))|_{b_{-6}}
\stackrel{m_3}{\rightarrow}
H^{1} (dP_9, {\cal O}_{dP_9} (6 \sigma|_{dP_9} -F))|_{c_{-3}}.
\label{64}
\end{equation}
Note, that
\begin{equation}
h^{1} (dP_9, {\cal O}_{dP_9} (3 \sigma|_{dP_9} -6F))|_{b_{-6}} =5
\label{65}
\end{equation}
and
\begin{equation}
h^{1} (dP_9, {\cal O}_{dP_9} (6 \sigma|_{dP_9} -F))|_{c_{-3}} =2.
\label{66}
\end{equation}
An explicit matrix for $m_3$ is most easily obtained if we now use the
Leray
spectral sequences and Serre duality discussed
in~\eqref{first},~\eqref{second}
and~\eqref{third},~\eqref{fourth} to identify
\begin{equation}
H^{1} (dP_9, {\cal O}_{dP_9} (3 \sigma|_{dP_9} -6F))|_{b_{-6}} =
H^{0} ({\cal S}, {\cal O}_{{\cal S}} (4))^{*}
\label{67}
\end{equation}
and
\begin{equation}
H^{1} (dP_9, {\cal O}_{dP_9} (6 \sigma|_{dP_9} -F))|_{c_{-3}}=
H^{0} ({\cal S}, {\cal O}_{{\cal S}} (1))^{*}.
\label{68}
\end{equation}
If we define the two dimensional linear space
\begin{equation}
\hat{V}= H^{0}({\cal S}, {\cal O}(1)),
\label{vec}
\end{equation}
then we see that
\begin{equation}
H^{1} (dP_9, {\cal O}_{dP_9} (3 \sigma|_{dP_9} -6F))|_{b_{-6}}=
Sym^{4}\hat{V}^{*}
\label{69}
\end{equation}
and
\begin{equation}
H^{1} (dP_9, {\cal O}_{dP_9} (6 \sigma|_{dP_9} -F))|_{c_{-3}}=
\hat{V}^{*},
\label{70}
\end{equation}
where by $Sym^{k}\hat{V}^{*}$ we denote the $k$-th symmetrized tensor
product
of the dual vector space $\hat{V}^{*}$ of $\hat{V}$.
Similarly, it follows from~\eqref{mod1} and~\eqref{vec} that $m_3$
is an element in
\begin{equation}
H^{0}(dP_9, {\cal O}_{dP_9} (3 \sigma|_{dP_9} +5F))|_{m_3} =
Sym^{3}\hat{V}.
\label{71}
\end{equation}
Let us now introduce a basis
\begin{equation}
\{u, v\} \in \hat{V}
\label{72}
\end{equation}
and the dual basis
\begin{equation}
\{u^{*}, v^{*}\} \in \hat{V}^{*},
\label{73}
\end{equation}
where
\begin{equation}
u^{*} u =v^{*} v =1, \qquad u^{*}v =v^{*} u =0.
\label{74}
\end{equation}
Then the space $Sym^{k}\hat{V}^{*}$ is spanned by all possible homogeneous
polynomials in $u^*, v^{*}$ of degree $k$. Specifically,
\begin{equation}
\{u^{*4}, u^{*3}v^{*}, u^{*2}v^{*2}, u^{*}v^{*3}, v^{*4} \}\in
Sym^{4}\hat{V}^{*}
\label{75}
\end{equation}
is a basis of $Sym^{4}\hat{V}^{*}$ and
\begin{equation}
\{u^{3}, u^{2}v, uv^{2}, v^{3} \}\in Sym^{3}\hat{V}.
\label{s3}
\end{equation}
is a basis of $Sym^{3}\hat{V}$.
Clearly, any section $m_3$ can be written in the basis~\eqref{s3} as
\begin{equation}
m_3 = \phi_1 u^3 +\phi_2 u^2v + \phi_3 uv^2 + \phi_4 v^3,
\label{76}
\end{equation}
where $\phi_{a}, a=1, \cdots 4$ represent the associated moduli. Now, by
using
the multiplication rules~\eqref{74}, we find that the explicit $2 \times
5$
matrix representation of $m_3$ in this basis, that is, the
$M_{11}$ submatrix of $M_{IJ}$, is given by
\begin{equation}
\bordermatrix{    & u^{*4}  & u^{*3}v^{*} & u^{*2}v^{*2}  & u^{*}v^{*3} &
v^{*4} \cr
              u^* & \phi_1  & \phi_2      & \phi_3        & \phi_4      &
0      \cr
              v^* & 0       & \phi_1      & \phi_2        & \phi_3      &
\phi_4  \cr}.
\label{77}
\end{equation}
Continuing in this manner, we can fill out the complete $20 \times 20$
matrix $M_{IJ}$.
It is not particularly enlightening, so we will not present the matrix
$M_{IJ}$
in this paper. What is important is the determinant of $M_{IJ}$. Let us
parametrize the
sections $m_2$ and $m_5$ as
\begin{eqnarray}
&&m_2 = \chi_1 u^2 + \chi_2 uv + \chi_3 v^2 \nonumber \\
&&m_5 = \psi_1 u^5 + \psi_2 u^4v + \psi_3 u^3v^2 + \psi_4 u^2v^3
+ \psi_5 uv^4 + \psi_6 v^5
\label{78}
\end{eqnarray}
where $\chi_b, b=1, 2, 3$ and $\psi_c, c=1, \cdots 6$ represent the
associated
moduli. It is then straightforward to compute the determinant of
$M_{IJ}$ using the programs Maple or MATHEMATICA. We find that
it has two rather amazing prperties: It factors very simply,
and it is independent of some of the variables. Namely
\begin{equation}
det{\ }f_{{\cal{C}}|_{dP_{9}}} = det{\ }M_{IJ} = {\cal{P}}^{4},
\label{79}
\end{equation}
where
\begin{eqnarray}
&&{\cal{P}} =
\chi_1^2 \chi_3 \phi_3^2 -
\chi_1^2 \chi_2 \phi_3 \phi_4 -
2\chi_1 \chi_3^2  \phi_3 \phi_1 - \nonumber \\
&&\chi_1 \chi_2 \chi_3  \phi_3 \phi_2 +
\chi_2^2 \chi_3  \phi_1 \phi_3 +
\phi_4^2 \chi_1^3 -              \nonumber \\
&&2 \phi_2 \phi_4 \chi_3 \chi_1^2  +
\chi_1 \chi_3^2 \phi_2^2 +
3 \phi_1 \phi_4 \chi_1 \chi_2 \chi_3 + \nonumber \\
&&\phi_2 \chi_1 \phi_4 \chi_2^2 +
\phi_1^2 \chi_3^3 -
\phi_2 \chi_2 \phi_1 \chi_3^2-
\phi_4 \phi_1 \chi_2^3
\label{80}
\end{eqnarray}
is a homogeneous polynomial of degree 5 in
the seven transition moduli
$\phi_a$ and $\chi_b$. Note that none of the remaining six moduli $\psi_a$
appear in ${\cal{P}}$.
We will see in~\cite{BDOnew} that the quintic polynomial ${\cal{P}}$ has
a simple interpretation, as a resultant. This leads to a geometric description of
the hypersurface $D_{{\cal{P}}} \subset {\mathbb P}^{12}$ given by ${\cal{P}}=0$:
There is a smooth
eleven-dimensional variety, ${\cal{F}}$, which is a ${\mathbb P}^{9}$ bundle over
${\mathbb P}^{1} \times {\mathbb P}^{1}$, and a map $i:{\cal{F}}
\rightarrow {\mathbb P}^{12}$ with the property that $i$ embeds each
${\mathbb P}^{9}$ fiber as a linear subspace ${\mathbb P}^{9}$ in ${\mathbb
P}^{12}$. Then, we find that
\begin{equation}
D_{{\cal{P}}}=i({\cal{F}})
\label{82aa}
\end{equation}
and, hence, $D_{{\cal{P}}}$ is the union of a two-parameter family of linear
subspaces ${\mathbb P}^{9}$.  The singular locus of $D_{{\cal{P}}}$
is the image of the loci of ${\cal{F}}$ on which $i$ is not
injective. These singular subspaces of $D_{{\cal{P}}}$ can be analyzed
completely. The fact that ${\cal{P}}$ does not depend on the $\psi$
coordinates means that the projective subspace
${\mathbb P}^{5} \subset {\mathbb P}^{12}$ given by
$\phi=\chi=0$ is contained in each of the subspaces ${\mathbb P}^{9}$, and hence
${\mathbb P}^{5}$ is contained in the singular locus of $D_{{\cal{P}}}$. In fact,
$D_{{\cal{P}}}$ is a cone whose vertex is this ${\mathbb P}^{5}$ and whose base is
a hypersurface $\overline{D_{{\cal{P}}}}$ in the complementary (quotient)
projective space ${\mathbb P}^{6}$ with homogeneous coordinates $\phi$ and $\chi$.

\section{The Superpotential and its Critical Points:}

In the previous section, we categorized the vanishing locus of ${\cal{P}}$
and,
hence, Pfaff$({\cal{D}}_{-})$. However, one can achieve much more than
this,
actually calculating from the above results the exact expressions for the
Pfaffian and the non-perturbative superpotential $W$. Recall that
${\cal{P}}$ is a section of ${\cal O}_{{\mathbb P}^{12}}(D_{{\cal{P}}})$
which vanishes on $D_{{\cal{P}}} \subset {\mathbb P}^{12}$.
On the other hand, Pfaff$({\cal{D}}_{-})$ is itself a
global holomorphic section of a line bundle over ${\mathbb P}^{12}$. That
it is a section, rather than a function, is a reflection of the fact that
the Pfaffian is not gauge invariant. Since, from the above results,
Pfaff$({\cal{D}}_{-})$ also has $D_{{\cal{P}}}$ as its zero locus, it
follows
that Pfaff$({\cal{D}}_{-})$ is a section of
\begin{equation}
{\cal O}_{{\mathbb P}^{12}}(pD_{{\cal{P}}}),
\label{83}
\end{equation}
where $p$ is a positive integer. Therefore,
\begin{equation}
{\rm Pfaff}({\cal{D}}_{-})=c{\cal{P}}^{p}
\label{83a}
\end{equation}
for some constant parameter $c$. 
It is possible to define a purely algebraic analogue of
${\rm Pfaff}({\cal{D}}_{-})$. A more careful analysis of our argument shows
that, quite generally, the algebraic version of
${\rm Pfaff}({\cal{D}}_{-})$ equals 
$detf_{{\cal{C}}|_{dP_{9}}}$, up to a constant. Furthermore, it was shown
in \cite{Bismut1,Bismut2,Bismut3} that the algebraic and analytic notions 
of ${\rm Pfaff}({\cal{D}}_{-})$ agree. Therefore,
\begin{equation}
{\rm Pfaff}({\cal{D}}_{-})=cdetf_{{\cal{C}}|_{dP_{9}}}=
c{\cal{P}}^{4},
\label{83.b}
\end{equation}
where ${\cal{P}}$ is given in~\eqref{80}. So in our specific case 
\begin{equation}
p=4.
\label{85}
\end{equation}
Thus, up to an overall
constant, we have determined Pfaff$({\cal{D}}_{-})$ as an explicit
holomorphic
function of the twelve moduli of ${\mathbb P}H^{0}
({\cal S}, {\cal O}_{{\cal S}}(5) \oplus {\cal O}_{{\cal S}}(3) \oplus
{\cal O}_{{\cal S}}(2))$. In addition, as will be discussed
in~\cite{BDOnew},
polynomial ${\cal{P}}^{4}$ is closely related to the theta function, $\Theta$,
whose natural variables, in this example, can be expressed in terms of the
vector bundle moduli.

We can now present the final answer for the vector bundle moduli
contribution
to the non-perturbative superpotential. Since the superpotential is
proportional to the Pfaffian, we conclude that
\begin{eqnarray}
&&W \propto {\cal{P}}^{4} =
(\chi_1^2 \chi_3 \phi_3^2 -
\chi_1^2 \chi_2 \phi_3 \phi_4 -
2\chi_1 \chi_3^2  \phi_3 \phi_1 - \nonumber \\
&&\chi_1 \chi_2 \chi_3  \phi_3 \phi_2 +
\chi_2^2 \chi_3  \phi_1 \phi_3 +
\phi_4^2 \chi_1^3 -              \nonumber \\
&&2 \phi_2 \phi_4 \chi_3 \chi_1^2  +
\chi_1 \chi_3^2 \phi_2^2 +
3 \phi_1 \phi_4 \chi_1 \chi_2 \chi_3 + \nonumber \\
&&\phi_2 \chi_1 \phi_4 \chi_2^2 +
\phi_1^2 \chi_3^3 -
\phi_2 \chi_2 \phi_1 \chi_3^2-
\phi_4 \phi_1 \chi_2^3)^{4} \; ,
\label{86}
\end{eqnarray}
where the thirteen transition moduli $\phi_{a}$, $\chi_{b}$ and $\psi_{c}$
parameterize the twelve dimensional moduli space ${\mathbb P}H^{0}
({\cal S}, {\cal O}_{{\cal S}}(5) \oplus {\cal O}_{{\cal S}}(3) \oplus
{\cal O}_{{\cal S}}(2))$. Summarizing, $W$ in~\eqref{86} is the
non-perturbative superpotential induced by wrapping a heterotic
superstring
once around the isolated curve ${\cal{S}}$ in an elliptically fibered
Calabi-Yau threefold with base $B={\mathbb F}_{1}$. The holomorphic vector
bundle $V$ has structure group $G=SU(3)$ and $W$ is a holomorphic function
of
the moduli associated with $V|_{{\cal{S}}}$. Note that, in this specific
case, the
$\psi_{c}$ moduli do not appear. This is an artifact
our our example. Generically, we expect all transition moduli to appear in
$W$.
The remaining moduli of $V$, that is, those not associated with
$V|_{{\cal{S}}}$, do not appear in this contribution to the
superpotential.

Having found an explicit expression for the non-perturbative
superpotential
$W$, it is of interest to find its critical points, that is, those points
or
submanifolds of the moduli space ${\mathbb P}H^{0}
({\cal S}, {\cal O}_{{\cal S}}(5) \oplus {\cal O}_{{\cal S}}(3) \oplus
{\cal O}_{{\cal S}}(2))$ where $N=1$ supersymmetry is preserved and the
cosmological constant vanishes. Recall (see
for example~\cite{WB}) that the Kahler covariant derivative of $W$ is
given by
\begin{equation}
D_{a_{i}}W=\partial_{a_{i}}W+\frac{1}{M_{P}^{2}}(\partial_{a_{i}}K)W,
\label{near}
\end{equation}
where $a_{i}$ is any moduli field in $W$, $K$ is the Kahler potential,
$M_{P}$
is the Planck mass and
that $N=1$ supersymmetry is unbroken if and only if
\begin{equation}
D_{a_{i}}W=0.
\label{near2}
\end{equation}
Using the expression
\begin{equation}
{\bf V
\rm}=e^{\frac{K}{M_{P}^{2}}}(G^{-1}|DW|^{2}-\frac{3}{M_{P}^{2}}|W|^{2})
\label{near3}
\end{equation}
for the potential energy, we see that both $DW$ and ${\bf V \rm}$ will
vanish
if and only if
\begin{equation}
W=dW=0.
\label{near4}
\end{equation}
These equations define the critical points of $W$.
To determine the critical points on a given coordinate patch
of ${\mathbb P}H^{0}
({\cal S}, {\cal O}_{{\cal S}}(5) \oplus {\cal O}_{{\cal S}}(3) \oplus
{\cal O}_{{\cal S}}(2))$, one performs the
differentiations in~\eqref{near4} with respect to homogeneous coordinates
and then goes to a local patch,
setting one of the homogeneous coordinates $\phi_a$, $\chi_b$ or $\psi_c$
to
unity. This proceedure can be justified using the Euler equation
for a homogeneous function. Since the superpotential $W$ is given
by the fourth power of the polynomial ${\cal P}$,
the solution to
the equations~\eqref{near4} is the whole zero locus of ${\cal P}$, that is,
the eleven-dimensional submanifold
\begin{equation}
D_{\cal P}
\subset {\mathbb P}H^{0}
({\cal S}, {\cal O}_{{\cal S}}(5) \oplus {\cal O}_{{\cal S}}(3) \oplus
{\cal O}_{{\cal S}}(2)).
\label{final1}
\end{equation}
The singularities of $D_{\cal P}$ are therefore irrelevant in our specific case.
In~\cite{BDOnew} we will study other examples, where some of the factors of
$W$ occur with multiplicity $1$, so a detailed geometric understanding of
$D_{\cal P}$ and its singularities becomes important.


\section{Conclusion:}


In this paper, we have considered non-perturbative superpotentials $W$
generated by wrapping a heterotic superstring once around an isolated
holomorphic curve $C$ of genus-zero in an elliptically fibered Calabi-Yau
threefold with holomorphic vector bundle $V$. We presented a method for
calculating the Pfaffian factor in such superpotentials as an explicit
function of the vector bundle moduli associated with $V|_{C}$. For
specificity, the vector bundle moduli contribution to $W$ was computed
exactly
for a Calabi-Yau manifold with base $B={\mathbb F}_{1}$ and isolated curve
${\cal{S}}$, and the associated critical points discussed. Our method,
however, has wide applicability, as will be shown in~\cite{BDOnew} where
the
vector bundle moduli contributions to the superpotentials in a number of
different contexts will be exactly computed and analyzed. In addition, we
will
show in~\cite{BDOnew} how to compute some of the homogeneous polynomials
appearing in these superpotentials analytically and will further extend
our
results to non-isolated holomorphic curves. Finally, in conjunction with
the
associated Kahler potential, one can use our superpotential to calculate
the potential energy functions of the vector bundle moduli. This potential
determines the stability of the vector bundle and has important
implications
for superstring and $M$-theory cosmology, as will be discussed elsewhere.



\end{document}